\documentclass[aps,prl,reprint,superscriptaddress]{revtex4-2}
\usepackage{graphicx}% Include figure files
\usepackage{dcolumn}% Align table columns on decimal point
\usepackage{bm}% bold math
\usepackage{color}
\usepackage{braket}
\usepackage{amsmath}

\newcommand{\beginsupplement}{
    \setcounter{table}{0}
    \renewcommand{\thetable}{S\arabic{table}}
    \setcounter{figure}{0}
    \renewcommand{\thefigure}{S\arabic{figure}}
}

\begin{document}

%\title{Quantum Computing Experiment on Electronic Band Structures}
\title{Demonstrating Quantum Computation for Quasiparticle Band Structures}

\author{Takahiro Ohgoe}
\affiliation{Panasonic Holdings Corporation, 1006 Kadoma, Kadoma City, Osaka 571-8508, Japan}

\author{Hokuto Iwakiri}
\affiliation{QunaSys Inc., Aqua Hakusan Building 9F, 1-13-7 Hakusan, Bunkyo, Tokyo 113-0001, Japan}

\author{Masaya Kohda}
\affiliation{QunaSys Inc., Aqua Hakusan Building 9F, 1-13-7 Hakusan, Bunkyo, Tokyo 113-0001, Japan}

\author{Kazuhide Ichikawa}
\affiliation{Panasonic Holdings Corporation, 1006 Kadoma, Kadoma City, Osaka 571-8508, Japan}

\author{Yuya O. Nakagawa}
\affiliation{QunaSys Inc., Aqua Hakusan Building 9F, 1-13-7 Hakusan, Bunkyo, Tokyo 113-0001, Japan}

\author{Hubert Okadome Valencia}
\affiliation{QunaSys Inc., Aqua Hakusan Building 9F, 1-13-7 Hakusan, Bunkyo, Tokyo 113-0001, Japan}

\author{Sho Koh}
\affiliation{QunaSys Inc., Aqua Hakusan Building 9F, 1-13-7 Hakusan, Bunkyo, Tokyo 113-0001, Japan}

\date{\today}

\begin{abstract}

Understanding and predicting the properties of solid-state materials from first-principles has been a great challenge for decades. Owing to the recent advances in quantum technologies, quantum computations offer a promising way to achieve this goal. Here, we demonstrate the first-principles calculation of a quasiparticle band structure on actual quantum computers. This is achieved by hybrid quantum-classical algorithms in conjunction with qubit-reduction and error-mitigation techniques. Our demonstration will pave the way to practical applications of quantum computers.
\end{abstract}

\maketitle

{\it Introduction}.---
Simulating solid-state materials accurately has been a long-standing issue. A widely used approach relies on first-principles calculations based on the density functional theory (DFT)~\cite{hohenberg1964,kohn1965}. Despite its success, DFT fails to describe accurately the electron correlation, and band gaps for semiconductors and insulators are systematically underestimated. Many body perturbation theories such as the GW approximation overcomes this problem partially~\cite{hybertsen1985}, while strongly correlated systems are still not properly described. Recently, wavefunction-based theories originally developed in the field of quantum chemistry have been successfully applied to solids~\cite{mcclain2017,dittmer2019,wang2020,gallo2021}. Moreover, the equation-of-motion coupled-cluster theory with single, double and triple excitations (EOM-CCSDT) is arguably more accurate than the GW approximation~\cite{lange2018}. Nevertheless, the high computational costs of such calculations prohibit practical applications to solid-state materials.

Owing to the recent advances in quantum technologies, quantum computations offer a promising way to efficient and accurate simulations of quantum systems. The two leading quantum algorithms for solving electronic-structure problems are the quantum phase estimation (QPE)~\cite{kitaev1995quantum,cleve1998,guzik2005} and the variational quantum eigensolver (VQE)~\cite{peruzzo2014}. QPE is expected to provide an exponential speedup in exact full configuration interaction (FCI) calculations. However, QPE requires a quantum circuit with many gates, and thus it is sensitive to the noise on quantum devices, hindering its practical applications until the realization of large-scale fault-tolerant quantum computers. On the other hand, VQE can be implemented with a shallower quantum circuit, allowing its application on near-term noisy quantum devices as demonstrated for small molecules~\cite{peruzzo2014,kandala2017,kandala2019}. There is therefore a scenario in which the practical application of quantum computers might be realized even in the present noisy intermediate-scale quantum (NISQ) era~\cite{preskill2018}.

Recently, VQE has been extended to periodic materials~\cite{liu2020,cerasoli2020,mizuta2021,fan2021,yoshioka2022,manrique2021,yamamoto2022,sherbert2021}.
% In particular, the quasiparticle band structures are also obtained by using a quantum subspace expansion (QSE) type formalism, which is based on EOM approach\cite{fan2021,yoshioka2022} with the projector operator method\cite{Szekeres2001}.
Interestingly, the quasiparticle band structures can now be obtained using the quantum subspace expansion (QSE)-type formalisms~\cite{mcclean2017,fan2021,yoshioka2022}, which are variants of EOM.
So far, experimental demonstrations of band structures on quantum devices are limited to the tight-binding (one-body) Hamiltonian~\cite{cerasoli2020,sherbert2021}. However, the inclusion of two-body terms is crucial to treat electron correlations, and it is indispensable for practical first-principles calculations.

In this paper, we experimentally demonstrate the first-principles calculation of a quasiparticle band structure on noisy quantum devices. We treat an {\it ab\ initio} Hamiltonian including the two-body terms. Calculations are based on QSE in conjunction with VQE. To mitigate noise effects on calculation results, we use the qubit-reduction and error-mitigation techniques. Owing to this methodology, we successfully reproduce the ideal noise-free results. Our demonstration is an important step toward practical band-structure calculation on quantum computers.

{\it Ab initio Hamiltonian}.---
We consider the second quantized representation of {\it ab\ initio} Hamiltonians of periodic systems which is given by
\begin{eqnarray}
  {\hat H} & = & \sum_{pq} \sum_{\mathbf k} t_{pq}^{\mathbf k} {\hat c}_{p{\mathbf k}}^{\dagger}  {\hat c}_{q{\mathbf k}}^{\vphantom{\dagger}}  \nonumber  \\
             &    &  +  \sum_{pqrs}  \sum_{{\mathbf k}_p {\mathbf k}_q {\mathbf k}_r {\mathbf k}_s}^{\prime}  v_{pqrs}^{{\mathbf k}_p {\mathbf k}_q {\mathbf k}_r {\mathbf k}_s} {\hat c}_{p{\mathbf k}_{p}}^{\dagger} {\hat c}_{q{\mathbf k}_{q}}^{\dagger} {\hat c}_{r{\mathbf k}_{r}}^{\vphantom{\dagger}} {\hat c}_{s{\mathbf k}_{s}}^{\vphantom{\dagger}}.
\end{eqnarray}
Here, ${\hat c}_{p{\mathbf k}}^{\dagger}$ (${\hat c}_{p{\mathbf k}}^{\vphantom{\dagger}} $) is the fermionic creation (annihilation) operator on the $p$th crystalline orbital (CO) with the crystal momentum ${\mathbf k}$ obtained through the Hartree-Fock (HF) calculation. The complex coefficients $t_{pq}^{\mathbf k} $ and $v_{pqrs}^{{\mathbf k}_p {\mathbf k}_q {\mathbf k}_r {\mathbf k}_s}$ are one-body and two-body integrals between COs, respectively. The prime on the summation symbol accounts for the momentum conservation law ${\mathbf k}_p + {\mathbf k}_q - {\mathbf k}_r - {\mathbf k}_s = {\mathbf G}$, where ${\mathbf G}$ is a reciprocal lattice vector of the unit cell.

{\it Hybrid quantum-classical algorithms}.---
To calculate quasiparticle band structures, we use the QSE method combined with VQE as reported in Refs.~\cite{fan2021,yoshioka2022}. In QSE, the ground state is used as a reference state $| \psi \rangle$. With an excitation operator ${\hat O}_{i}$, we construct a low-energy subspace Hamiltonian ${\mathbf H}^{\rm sub}$ which is given by $H_{ij}^{\rm sub} = \langle \psi | {\hat O}_{i}^{\dagger} {\hat H} {\hat O}_{j}^{\vphantom{\dagger}} | \psi \rangle$ in the matrix representation. Then we obtain the spectrum of energy eigenvalues (quasiparticle bands) by solving the generalized eigenvalue problem ${\mathbf H}^{\rm sub} {\mathbf C} = {\mathbf S}^{\rm sub} {\mathbf C} {\mathbf E}$. Here, ${\mathbf C}$ is the matrix consisting of the eigenvectors and ${\mathbf E}$ is the diagonal matrix whose elements are eigenvalues. In addition, the matrix ${\mathbf S}^{\rm sub}$ is given by $S_{ij}^{\rm sub} = \langle \psi | {\hat O}_{i}^{\dagger} {\hat O}_{j}^{\vphantom{\dagger}} | \psi \rangle$. We evaluate the matrix elements $H_{ij}^{\rm sub}$ and $S_{ij}^{\rm sub}$ using quantum devices. The diagonalization for solving the generalized eigenvalue problem is performed on classical computers. As the excitation operator ${\hat O}_{l {\mathbf k}}^{\vphantom{\dagger}}$, we choose ${\hat c}_{l {\mathbf k}}^{\vphantom{\dagger}}$ for obtaining the valence bands at each ${\mathbf k}$, where $l$ runs over the occupied orbitals. Similarly, ${\hat c}_{l {\mathbf k}}^{\dagger}$ is used for the conduction bands, where $l$ runs over the unoccupied orbitals.

To prepare the ground state $| \psi \rangle$, we use the VQE method which is based on the variational method with a parameterized wavefunction ansatz. The wavefunction ansatz takes the form of $|\psi ({\bm \theta}) \rangle = {\hat U} ({\bm \theta}) \ket{\psi_0}$, and the parameters ${\bm \theta}$
are variationally optimized to minimize the energy expectation value. Here, $\ket{\psi_0}$ is an input quantum state and ${\hat U} ({\bm \theta})$ is a unitary operator implemented on a quantum circuit.

\begin{figure}[t]
\includegraphics[width=8cm]{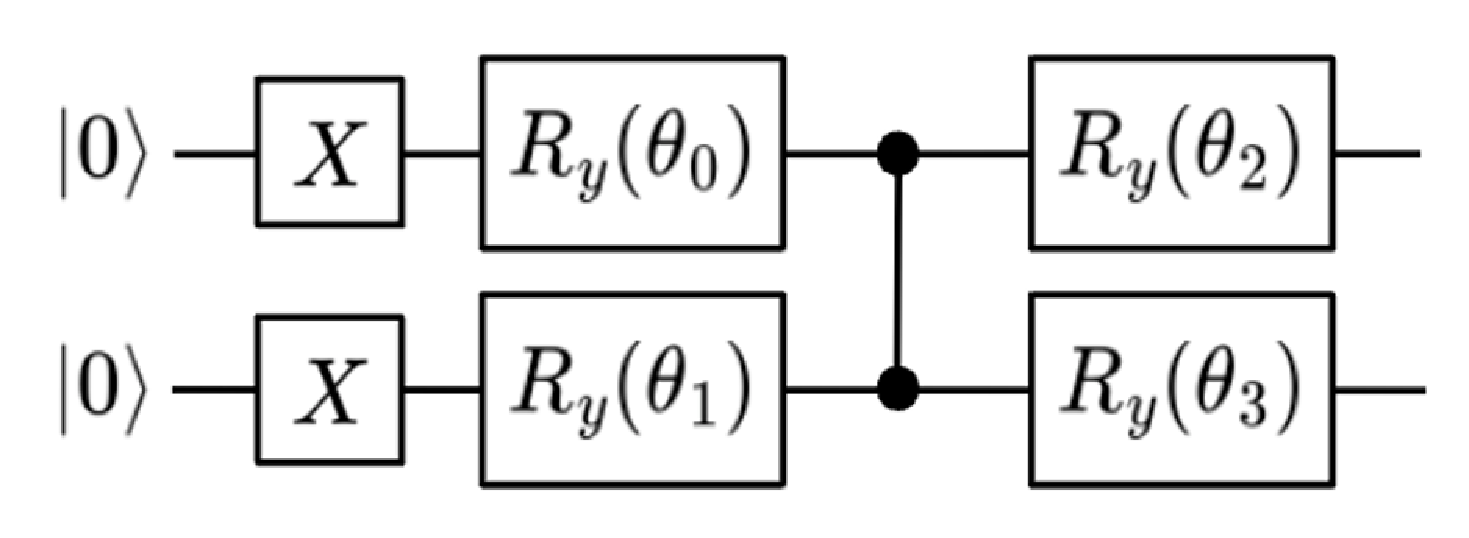}
\centering
\caption{\label{fig:ansatz} Quantum circuit used in the present study. The input state $\ket{\psi_0}$ and the unitary operator ${\hat U} ({\bm \theta})$ correspond to $\ket{00}$ and the whole of the operators, respectively. The HF state is $\ket{11}$ which we obtain after the operation of X-gates. Four variational parameters $\theta_i$ for the $R_y$ rotation gates are optimized during VQE calculations.}
\end{figure}

{\it Settings}.---
We consider Si that has the diamond crystal structure with the experimental lattice constant of 5.43 \AA, where two atoms are contained in the primitive cell. For the Brillouin zone sampling, a 1$\times$1$\times$1 k-point grid is centered at the target k-point on the band path. For each target k-point, the Hamiltonian is constructed separately. The GTH (Goedecker-Teter-Hutter) pseudpotential~\cite{goedecker1996} is used together with the GTH-SZV (single-zeta valence) basis set. To reduce the number of required qubits, we use the active space approximation and take the highest occupied and lowest unoccupied COs into account for each k-point. In addition, we map the target fermionic Hamiltonians to the qubit ones with the technique reported in Ref.~\cite{bravyi2017}, which tapers off two qubits by exploiting $Z_2$ symmetries. Thus, the Si crystal is solved by a two-qubit system.

 In Fig. \ref{fig:ansatz}, we show the quantum circuit used in the present study. It is motivated by the hardware-efficient ansatz~\cite{kandala2017}, but we consider a simplified circuit with the minimal depth to reduce the computational cost and the effect of noise. The variational parameters are optimized by using the sequential minimal optimization~\cite{nakanishi2020} on a noiseless statevector simulator.

 Quantum devices used in our demonstration are Rigetti's Aspen-M processors\cite{aspenM3}. In particular, we adopt qubits with high fidelities which are shown in Table \ref{tab:fidelity} as examples. In the QSE calculations, we apply the readout-error mitigation (REM)~\cite{chen2019,maciejewski2020,bravyi2021,hamilton2020}. Moreover, to cancel out the bias coming from time-varying noise, we repeat the same calculation 40-70 times and average the outcomes (See also Supplemental Material~\cite{suppl}).

\begin{table}[t]
\begin{center}
\caption{\label{tab:fidelity} Fidelities of qubits (selected on Feb. 6, 2023). Here, two selected qubits are labeled as Q1 and Q2, respectively.}

\begin{tabular*}{1.0\linewidth}{@{\extracolsep{\fill}}cccc}
\hline \hline
 Qubits  & Fidelity (\%) & CZ gate fidelity (\%)  & Readout (\%) \rule{0pt}{2.6ex} \\ \hline
 Q1, Q2  & 99.90, 99.97  &  98.29 & 98.10, 99.60 \rule{0pt}{2.6ex} \\
\hline \hline
\end{tabular*}
\end{center}
\end{table}

\begin{figure}[t]
\includegraphics[width=8.8cm]{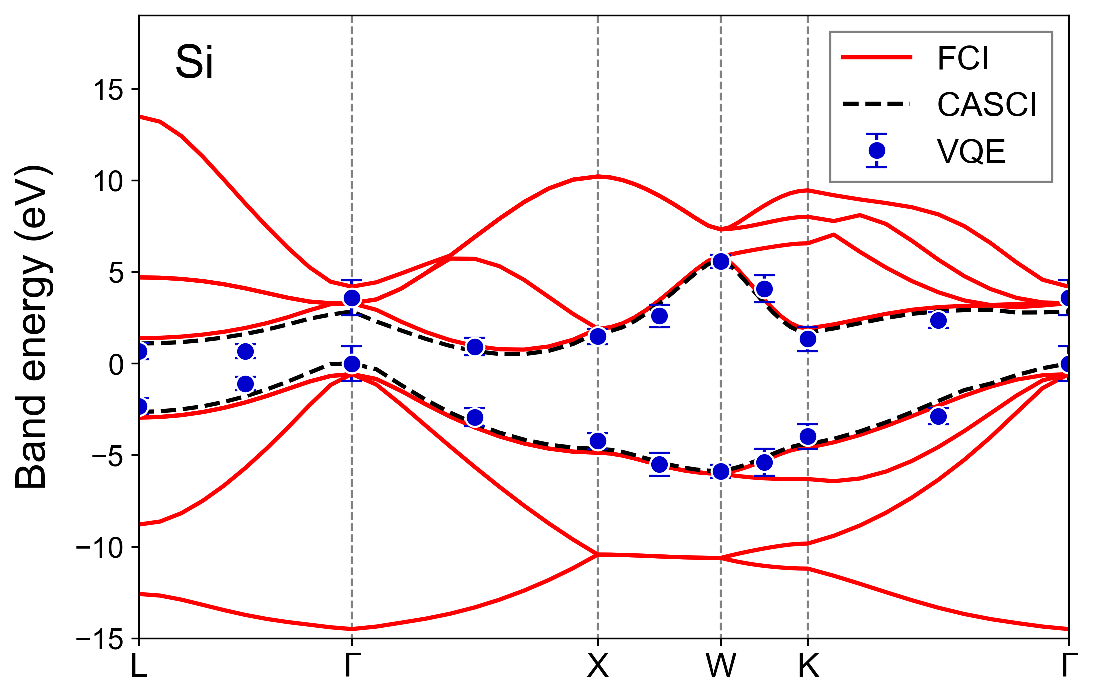}% Here is how to import EPS art
\caption{\label{fig:band} Quasiparticle band structure of the Si crystal obtained by QSE. Results with different ground states obtained by VQE, CASCI, and FCI are presented. The ground states were prepared on a classical computer. For the VQE ground states, the QSE calculations were performed on actual quantum devices. The valence band energy at the $\Gamma$ point obtained by CASCI is set to 0 eV.}
\end{figure}

{\it Results}.---
In Fig. \ref{fig:band}, we show the quasiparticle band structure obtained by QSE. The result labeled as ``VQE" represents the main one where the QSE calculations were performed on quantum devices. For comparisons, we include the results where the exact ground states were obtained using FCI or the complete active space configuration interaction (CASCI) with the same active space as VQE. For these two results, the QSE calculations were accurately performed using a statevector simulator. As seen in this figure, the VQE result agrees well with the corresponding FCI and CASCI ones, showing that the band structure is accurately obtained using actual quantum devices.

\begin{figure}[t]
\includegraphics[width=8.5cm]{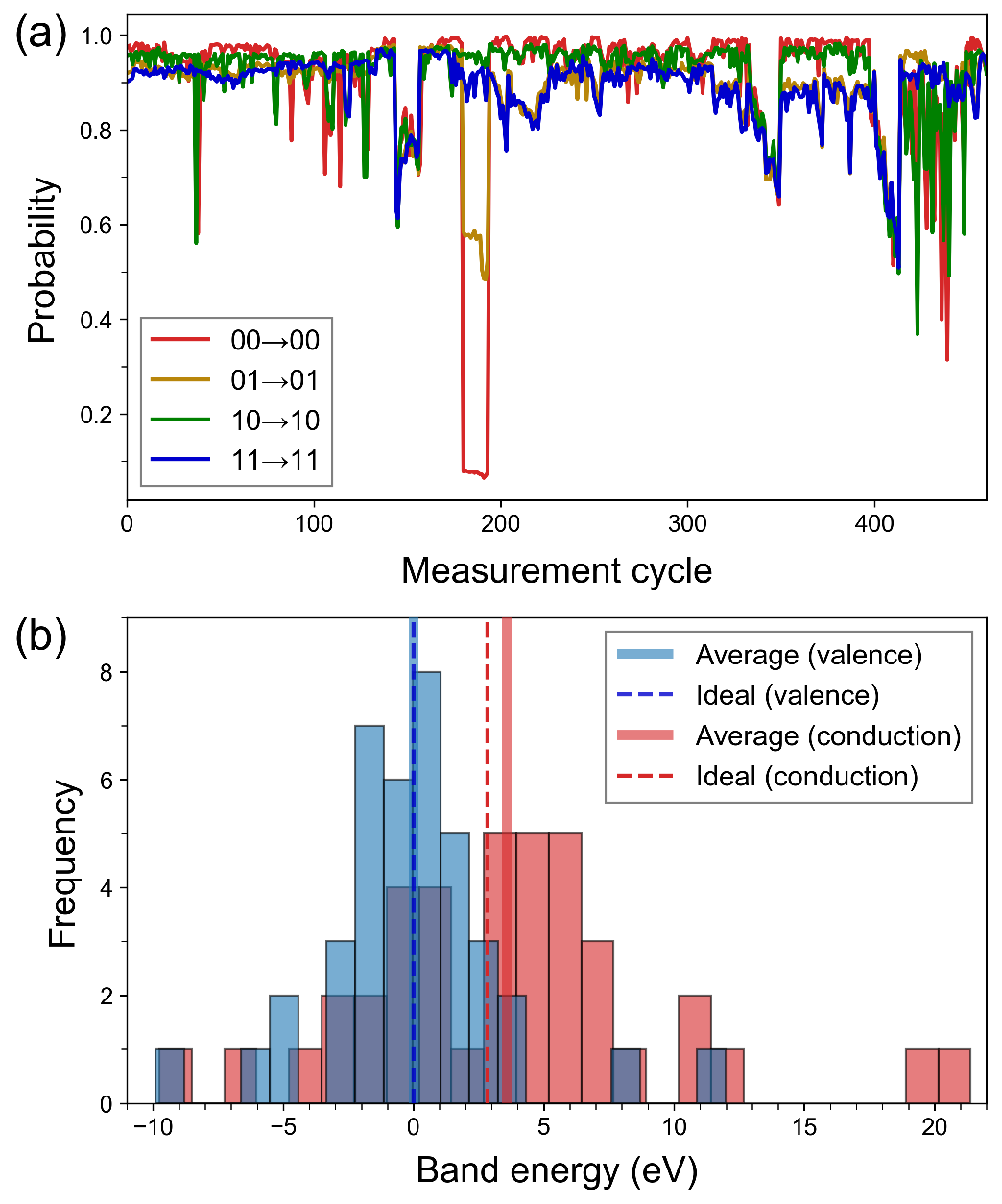}% Here is how to import EPS art
\caption{\label{fig:error} Error mitigation schemes. (a) Diagonal elements of the calibration matrix measured at various times (measurement cycle). (b) Histograms of 40 independent calculation results for the valence and conduction band energies at the $\Gamma$ point. In these results, REM was applied. The vertical dashed lines represent the ideal noise-free values obtained by a statevector simulator. On the other hand, the vertical solid lines show the mean values of the calculation results obtained on quantum devices. }
\end{figure}

To apply REM, we constructed a calibration matrix before each calculation. More specifically, we prepared all the computational basis states and measured each probability distribution in the same basis (See also the Supplemental Material for the detailed definition of the calibration matrix). In Fig. \ref{fig:error} (a), we show all the records of the calibration matrices (diagonal elements) measured throughout the present study (December 20, 2022 to March 27, 2023), indicating that the noise is time-varying and its effect on outcomes also varies at every measurement. Figure \ref{fig:error} (b) shows a histogram of the calculated band energies at the $\Gamma$ point. Even though REM is applied, most of the results still deviate from the ideal noise-free ones obtained by the statevector simulator. Since the statistical error in each calculation was much smaller than the deviation (See also Supplemental Material~\cite{suppl}), the bias is mainly due to the effect of device noise. However, we found that averaging the outcomes at various times leads to error cancellations, which explains our well-reproduced ideal values.

\begin{figure}[t]
\includegraphics[width=8.5cm]{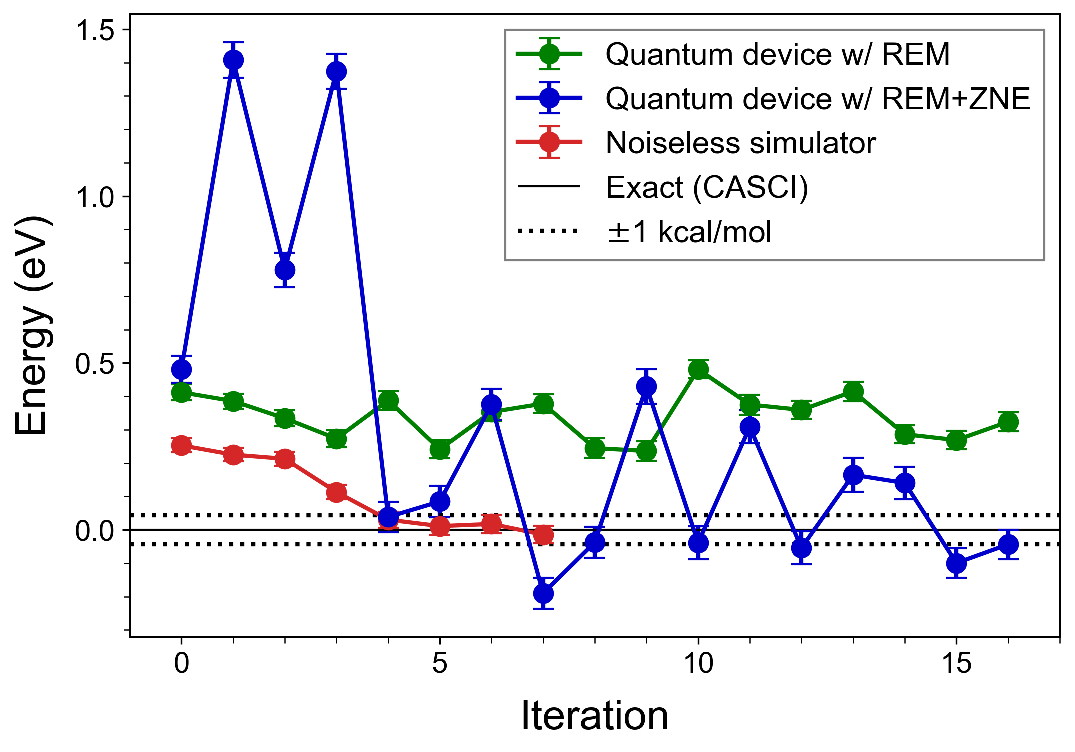}% Here is how to import EPS art
\caption{\label{fig:vqe} Energies in the VQE optimization processes obtained by a noiseless sampling simulator or actual quantum devices. In the latter, we applied REM or both REM and ZNE. The error bars represent statistical errors. The exact value by CASCI is set to 0 eV. The dashed horizontal lines denote the $\pm$ 1 kcal/mol precision compared with CASCI.}
\end{figure}

Finally we address the accuracy of the VQE ground state. Figure \ref{fig:vqe} shows optimization processes of the VQE state $|\psi ({\bm \theta}) \rangle$ at the $\Gamma$ point. The energy optimized by the noiseless sampling simulator agrees with the exact ground-state energy obtained by CASCI within the chemical accuracy of $\pm 1$ kcal/mol (= $\pm  0.0434$ eV), showing that the ansatz we employ describes the ground state accurately. For this particular k-point, we also performed the optimization using actual quantum devices as a demonstration. As a result, we found that the optimization with REM is not sufficient to reach the exact ground-state energy. To overcome it, we also applied the zero-noise extrapolation (ZNE) method~\cite{richardson1911,li2017,temme2017,kandala2019} at the cost of introducing the redundant gates and measurements  (See also Supplemental Material~\cite{suppl}). With this approach, we eventually observed a tendency of converging to the exact ground-sate energy.

{\it Summary and Outlook}.---
To summarize, we have demonstrated $\it ab\ initio$ calculations of a quasiparticle band structure using actual quantum devices. The difficulty arises from the noise inherent to quantum devices. To circumvent it, a qubit-reduction strategy combining the pseudopotential, the active space approximation, and the symmetry-based qubit tapering technique were implemented. Owing to this strategy, we could solve the Si crystal with a two-qubit system. Moreover, the error-mitigation techniques were successfully applied and the ideal noise-free results were reproduced. Since the band structure is an essential physical characterization of the periodic materials, we believe that the present demonstration is a crucial step towards practical applications of quantum computers to solid-state materials.

Scaling up the quantum computations is certainly an indispensable subsequent step. The REM method used in the present study requires manipulations with probability vectors of size $2^n$, which limits the scalability ($n$ is the number of qubits). Problems on large systems might be tackled with scalable error-mitigation methods~\cite{bravyi2021,hamilton2020,nation2021,srinivasan2022}. In addition, a promising alternative approach is the application of recently developed hybrid quantum-classical algorithms which circumvents the VQE optimization difficulties. For example, a quantum algorithm combined with a quantum Monte Carlo was developed and demonstrated for the ground-state calculations of the diamond solid using 16-qubit systems~\cite{huggins2022}. In another example, a hybrid algorithm called quantum-selected configuration interaction was shown to calculate excited states as well as ground states~\cite{kanno2023}. These are interesting directions toward practical applications of quantum computers to larger systems.

{\it Acknowledgements}.---
The authors thank Amazon Web Services for supporting this work through their Amazon Braket service. All computations in the present work are performed using the quantum-circuit simulation library Qulacs~\cite{suzuki2021} with QURI Parts~\cite{quri}, a library for developing quantum algorithms. The Hamiltonians are constructed by the PySCF~\cite{sun2020} and the OpenFermion~\cite{mcclean2020} software.

% Create the reference section using BibTeX:
\bibliography{reference.bib}

%\appendix

\newpage
\clearpage

{\LARGE Supplemental Material}
\beginsupplement

\section{Quantum devices}
In the present study, we used Rigetti's quantum devices via Amazon Braket~\cite{braket}. More specifically, first we used the Aspen-M-2 processor, and later we used its successor, Aspen-M-3 processors~\cite{aspenM3}. For QSE, we performed the same calculations 40 times for each k-point to take the average. After that, for k-points where we observed relatively large variances of outcomes, we obtained 30 more calculation results (This was done for the L point and the middle point between L and $\Gamma$).

\begin{figure*}[h]
    \centering
    \includegraphics[width=13cm]{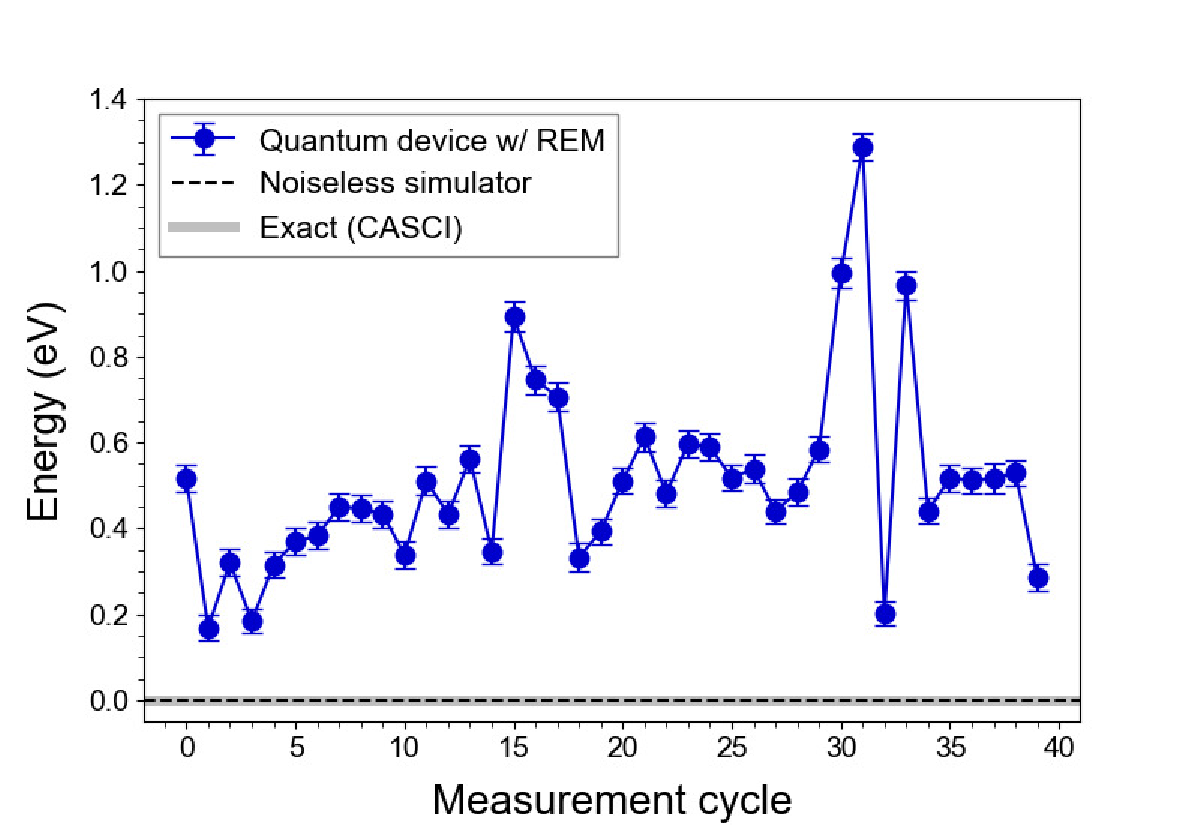}% Here is how to import EPS art
    \caption{Energies of the VQE optimized state at the $\Gamma$ point. We used the variational parameters fully optimized on a statevector simulator. Blue circles represent 40 independent results obtained on the quantum device over the course of 2 days. In these results, REM was applied. The error bars represent statistical errors. The dashed line denotes the ideal noise-free result obtained by the statevector simulator. The exact value by CASCI is set to 0 eV.}
    \label{fig:E_q_device}
\end{figure*}

Figure \ref{fig:E_q_device} shows calculation results for the energy of a VQE optimized state. The calculations were performed on the Aspen-M-2 processor and we obtained 40 independent results over the course of two days. As a result, we found that the amount of deviations from the ideal noise-free results obtained by the statevector simulator varies at each measurement. Since the statistical error in each calculation is much smaller than the deviation, it indicates that the noise is time-varying and the bias on outcomes also varies at each measurement. 

\section{Shot allocation}

In the QSE calculations, the matrix elements of ${\mathbf H}^{\rm sub}$ and ${\mathbf S}^{\rm sub}$ were evaluated on the quantum devices. More concretely, to estimate the expectation values for the reference state $\ket{\psi}$, 
the operators $\hat{O}_{i}^{\dagger} \hat{H} \hat{O}_{j}$ and $\hat{O}_{i}^{\dagger} {\hat O}_j $ were converted to the linear combinations of Pauli strings using the mapping technique reported in Ref. \cite{bravyi2017}.
We truncated the Pauli strings whose coefficients are below $10^{-8}$ in absolute values. In addition, we adopted the grouping method~\cite{kandala2017,verteletskyi2020} where qubit-wise commuting Pauli strings are categorized into the same group and they are measured simultaneously. In our calculations, we obtained one-to-five groups for $\hat{O}_{i}^{\dagger} \hat{H} \hat{O}_{j}$ and one group for $\hat{O}_{i}^{\dagger} {\hat O}_j $ for each k-point.
Then, 10,000 shots were allocated uniformly to each group. Here, the number of shots is the number of repetitions for the state preparation and measurement.

In the VQE calculations, the Hamiltonian was decomposed to five Pauli strings and grouped into two groups using the same qubit-wise commuting grouping method. To obtain the calculation results shown in Fig. 4 of the main text, we allocated 5,000 shots to each group.

\section{Readout-error mitigation}
The readout-error mitigation (REM)~\cite{chen2019,maciejewski2020,bravyi2021,hamilton2020} is a technique to mitigate noise effects that occur in the readout process. In this method, an ideal probability distribution ${\mathbf p}_{\rm ideal}$ of a quantum state is assumed to relate to the measured distribution ${\mathbf p}_{\rm noisy}$ as
\begin{equation}
    {\mathbf p}_{\rm noisy} = \mathbf{M} {\mathbf p}_{\rm ideal},
\end{equation}
where $\mathbf{M}$ is a calibration matrix. Based on this relation, we can infer ${\mathbf p}_{\rm ideal}$ from ${\mathbf p}_{\rm noisy}$ via the inverse transformation after obtaining $\mathbf{M}$ in the preliminary measurements. 

To obtain ${\mathbf M}$, we first prepared the computational basis states $\ket{00}$, $\ket{01}$, $\ket{10}$ and $\ket{11}$ separately, and performed measurements using 10,000 shots for each state. Then, the matrix ${\mathbf M}$ was constructed as
\begin{equation}
    {\mathbf M} = \frac{1}{N}
    \begin{pmatrix}
        M^{00}_{{00}} & M^{00}_{{01}} & M^{00}_{{10}} & M^{00}_{{11}} \\
        M^{01}_{{00}} & M^{01}_{{01}} & M^{01}_{{10}} & M^{01}_{{11}} \\
        M^{10}_{{00}} & M^{10}_{{01}} & M^{10}_{{10}} & M^{10}_{{11}} \\
        M^{11}_{{00}} & M^{11}_{{01}} & M^{11}_{{10}} & M^{11}_{{11}} 
    \end{pmatrix},
\end{equation}
where $M^{x}_{y}$ denotes the number of outcomes $x$ (= 00, 01, 10, 11) in the measurement of the $y$ basis state. $N$ is a normalization factor which is equal to 10,000 in the present study. 

\begin{figure*}[h]
    \centering
    \includegraphics[width=13cm]{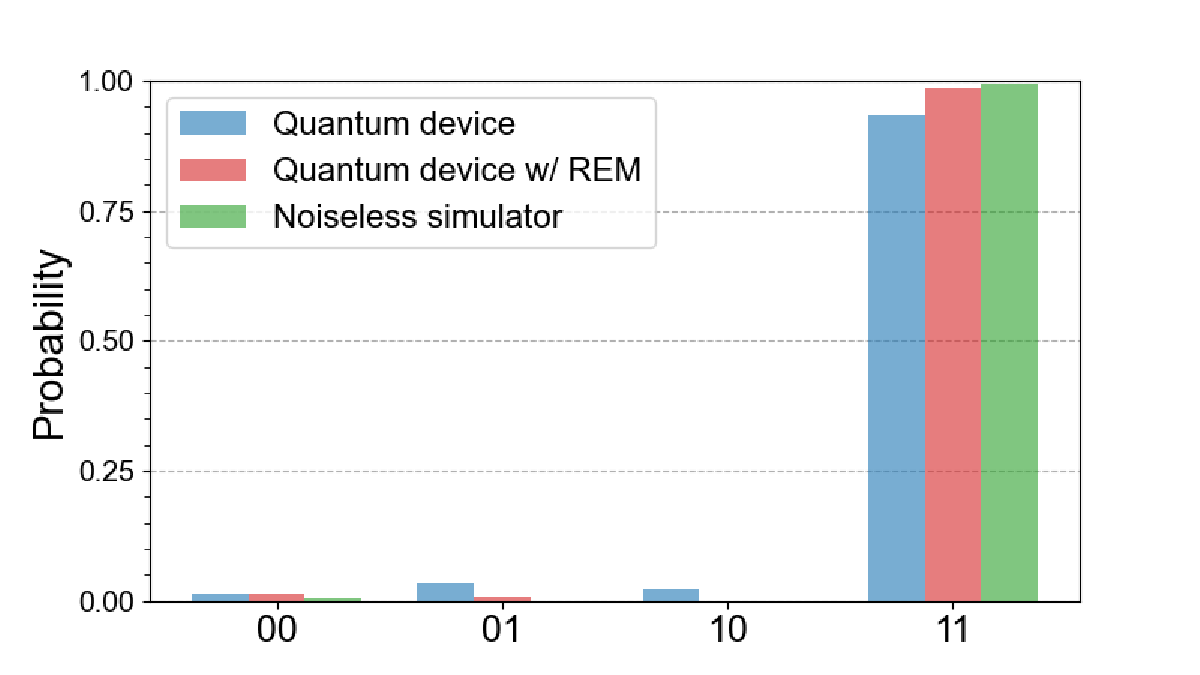}% Here is how to import EPS art
    \caption{Probability distributions of the VQE optimized state $| \psi \rangle$ at the L point. We used the variational parameters fully optimized on a statevector simulator. Blue, red, and green bars represent the results obtained by the quantum device, the quantum device with REM, and the noiseless simulator, respectively. }
    \label{fig:rem}
\end{figure*}

Figure \ref{fig:rem} shows an application result of REM. In this example, we measured the probability distribution of a VQE optimized state $| \psi \rangle$. From this result, we see that, owing to REM, we obtained the probability distribution which is close to the ideal one obtained by the noiseless simulator.

\section{Zero-noise extrapolation}

\begin{figure*}[h]
    \centering
    \includegraphics[width=18cm]{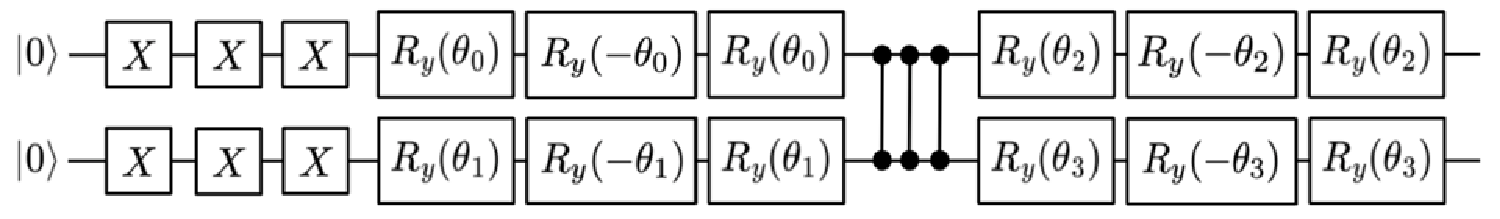}
    \caption{Quantum circuit corresponding to the noise-amplification factor $\lambda = 3$.}
    \label{fig:intentionally_noise_circuit}
\end{figure*}

\begin{figure*}[h]
    %\centering
    \includegraphics[width=13cm]{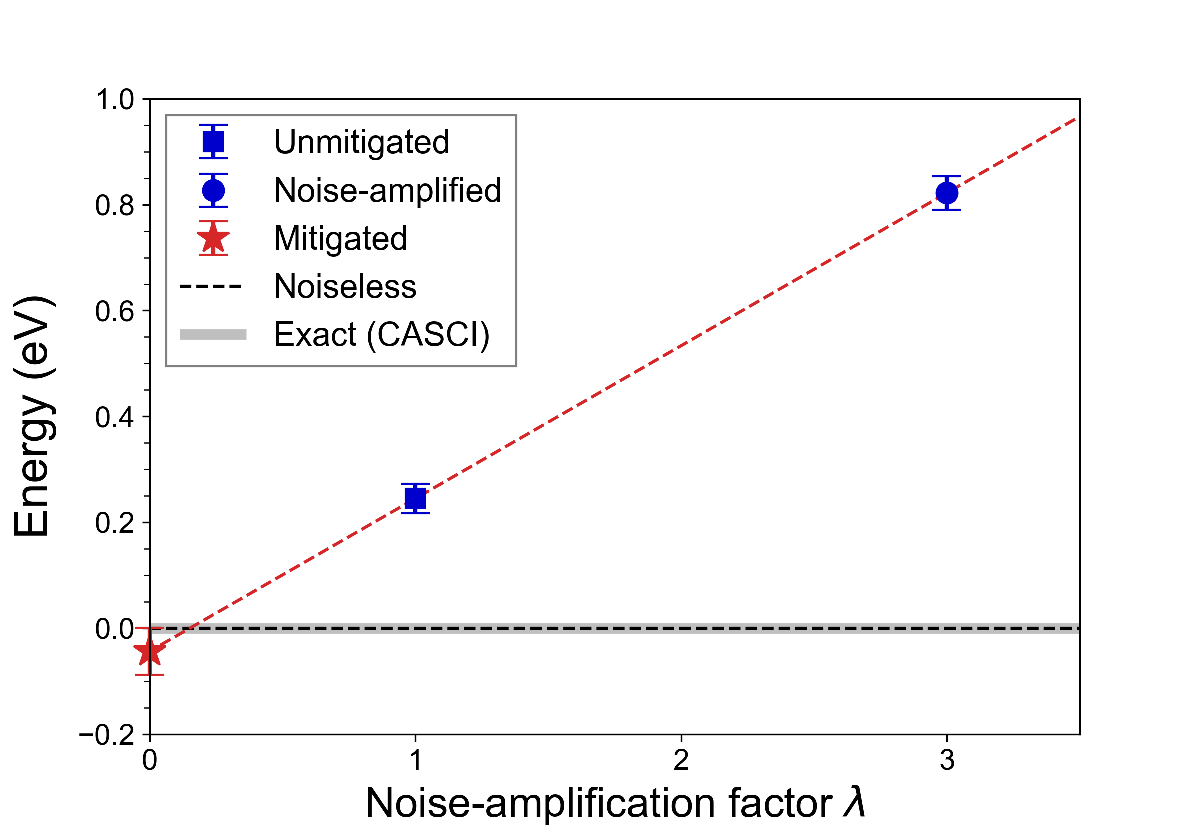}% Here is how to import EPS art
    \caption{Extrapolation of energies of a VQE optimized state at the $\Gamma$ point. The circle and square points represent the values obtained on the quantum device with REM for the original quantum circuit ($\lambda=1$) and the noise-amplified one  ($\lambda=3$), respectively. The star point denotes the extrapolated value at $\lambda=0$ (noiseless limit) obtained by the linear extrapolation. The exact value by CASCI is set to 0 eV.}
    \label{fig:E_q_noisy_simulator}
\end{figure*}

The zero-noise extrapolation (ZNE)~\cite{richardson1911,li2017,temme2017,kandala2019} is a technique for mitigating the effect of gate errors. In this approach, we estimate the noise-free result by extrapolation of outcomes with different noise levels. To this end, we prepared a noise-amplified quantum circuit by replacing each gate component $A$ with redundant one $A A^{\dagger} A$ as shown in Fig. \ref{fig:intentionally_noise_circuit}. We define a dimensionless noise-amplification factor $\lambda$ as the depth of this component. Namely, the original quantum circuit and the redundant one correspond to $\lambda = 1$ and 3, respectively. The extrapolations were performed in terms of $\lambda$ to estimate the value in the noiseless limit $\lambda=0$. Figure \ref{fig:E_q_noisy_simulator} shows an extrapolation result of the energy of a VQE optimized state.

\end{document}